\documentclass[a4paper,twoside]{article}

\usepackage{epsfig}
\usepackage{subcaption}
\usepackage{calc}
\usepackage{amssymb}
\usepackage{amstext}
\usepackage{amsmath}
\usepackage{amsthm}
\usepackage{multicol}
\usepackage{pslatex}
\usepackage{apalike}
\usepackage{algorithm2e}
\usepackage[bottom]{footmisc}

\usepackage{hyperref}
\usepackage[T1]{fontenc}
\usepackage[utf8]{inputenc}
\usepackage[english]{babel}
\usepackage{amsfonts}
\usepackage{graphicx}
\usepackage{color}
\usepackage{cleveref}
\usepackage{listings}

\usepackage{SCITEPRESS}     

\begin{document}

\title{Logic interpretations of ANN partition cells}


\author{\authorname{Ingo Schmitt\orcidAuthor{0000-0002-4375-8677}} 
\affiliation{Brandenburgische Technische Universität Cottbus-Senftenberg,  Germany}
\email{schmitt@b-tu.de}
}


\abstract{Consider a binary classification problem solved using a feed-forward artificial neural network (\texttt{ANN}). Let the \texttt{ANN} be composed of a ReLU layer and several linear layers (convolution, sum-pooling, or fully connected). We assume the network was trained with high accuracy.   Despite numerous suggested approaches, interpreting an artificial neural network remains challenging for humans.  For a new method of interpretation, we construct a bridge between a simple \texttt{ANN} and logic.   As a result, we can analyze and manipulate the semantics of an \texttt{ANN} using the powerful tool set of logic. To achieve this, we decompose the input space of the \texttt{ANN} into several network partition cells. Each network partition cell represents a linear combination that maps input values to a classifying output value.   For interpreting the linear map of a partition cell using logic expressions, we suggest minterm values as the input of a simple  \texttt{ANN}.   We derive logic expressions representing interaction patterns for separating objects classified as '1' from those classified as '0'. To facilitate an interpretation of logic expressions,  we present them as binary logic trees.}

\onecolumn \maketitle \normalsize \setcounter{footnote}{0} \vfill

\section{Introduction}
\label{sec:setting}

For relying on trained \texttt{ANN}s in critical applications, a human interpretation of their behavior is inevitable. An \texttt{ANN} is a numerical approach where unlike decision trees no navigation by Boolean decisions on single input attributes exists. Instead,  complex mappings hide internal decision logic, resulting in a lack of trust \cite{Mil19,KauUslRit22}. Therefore, an \texttt{ANN} is often called a black-box solution. A  task related to interpretation is to explain the classification result for a given input object. 

For example, the Shapley-value \cite{Sha53} and its adaptation SHAP \cite{Lun20} compute the importance of single input attributes as part of an additive measure. The SHAP approach generalizes the local interpretable model-agnostic explanations (LIME) \cite{Rib16}. Another proposal is to extract ridge and shape functions in the context of an extracted generalized additive model to illustrate the effect of attributes.  Approaches to measuring the importance of an input attribute can be generalized to measure the strength of interactions among several attributes, see for example \cite{MurSon93}.

Many researchers, see \cite{KraTschWei23,Mol20} for an overview, strive to make the black-box behavior more transparent and more comprehensible.


Note that decision tree classifiers  and their derivatives \cite{Sut16} are based on Boolean logic decisions. Logic expressions are seen as much easier to interpret than numerical classifiers like \texttt{ANN}s.  The work from \cite{Bal91,Gar92,Blu04} is  an early attempt to bridge neural networks (subsymbolic paradigm) with logic (symbolic paradigm). It defines schemata representing states of a neural network and fuzzy-logic-like operations (conjunction and disjunction).  The operations, however,  do not  obey the laws of the Boolean algebra.  A schema induces a partition of the input space that is essentially different from the partition proposed in our approach.

The work \cite{Gra04} discusses the concepts of 'incompatibility' and 'implementation' in the context of bridging the symbolic and the subsymbolic paradigms of information processing and relates both paradigms to the concepts of quantum mechanics.

An interesting approach \cite{Bal17,Yan18} to bridge logic in the form of decision trees with a deep neural network  is to simulate a decision tree using a neural network. After learning and combining the binnings of attribute values, linear networks are trained for every binning combination in order to learn the classification. In contrast, in our work, the starting point is a given trained simple \texttt{ANN},  which we attempt to interpret \emph{after} training.  Furthermore,   the partitioning of  \cite{Bal17,Yan18} is based on decision tree split nodes simulated by the softmax function whereas in our approach it is based on ReLU-nodes.


In our work, we utilize concepts from the theory of quantum logic \cite{Mit78} in order to bridge the gap between a trained \texttt{ANN}  and logic expressions \cite{schmitt2006quantum,schmitt2008qql,schmitt2013logic}. Conceptually, the evaluation of an input object against an \texttt{ANN} corresponds to a quantum measurement of a state vector against a projector, where commuting projectors form a Boolean sublattice of a quantum logic (orthomodular lattice). The quantum logic concepts used for evaluating logic expressions encompass concepts from probability theory of complex events and obey the laws of Boolean algebra. For an intuitive understanding in the context of an \texttt{ANN}, we explain the concepts of our approach by exclusive use of probability theory.



In addressing a binary classification problem, we begin with a set $O={(x_i,y_i)}$ of input-output pairs. A subset $TR\subseteq O$ represents the training data, where $x_i\in \mathbb{R}^n$ and $y_i\in \{0,1\}$. For classification, we utilize a feed-forward artificial neural network (\texttt{ANN)} composed of a ReLU layer and  several linear layers, such as sum-pooling, convolution, or fully connected layers. The output layer contains exactly one output node. Employing an output threshold operator $th_\tau(o)$ on the output value $o$ determines   the class decision:
$$\hat{y}:=th_\tau(o):=\left\{\begin{array}{ll}1&\text{if }o>\tau\\0&\text{otherwise.}\end{array}\right.$$
We restrict ourselves to a \emph{simple} \texttt{ANN}, that is,  the number  $n$ of attributes of an input object $x$ is small. 
We regard the input data as so-called \emph{tabular data}, where a human  can interpret every single input attribute.
Furthermore, we assume that the \texttt{ANN} was trained using $TR$ and a loss function (e.g. MSE) with high accuracy. In the sequel,  the training  of the \texttt{ANN} will play no significant role.

\section{Proposed Method in a Nutshell}

\begin{figure}
\begin{center}
\includegraphics[scale=0.3]{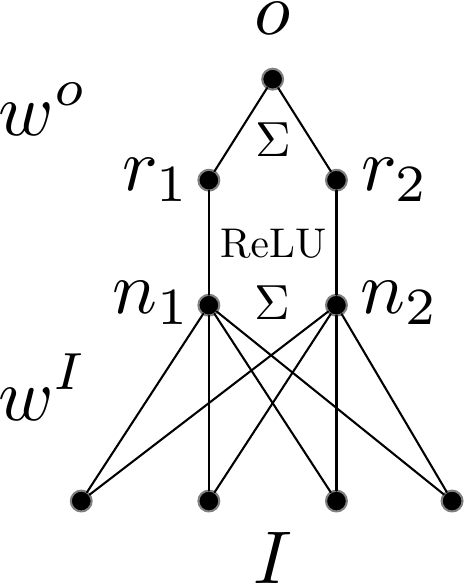}
\end{center}
\caption{\label{fig:simple_ann}Example of a simple \texttt{ANN}}
\end{figure}

For demonstrating a small \texttt{ANN},  Figure~\ref{fig:simple_ann} visualizes a very small example network with two ReLU nodes $r:=\begin{pmatrix}r_1&r_2\end{pmatrix}^t=\begin{pmatrix}ReLU(n_1)&ReLU(n_2)\end{pmatrix}^t$ with
 \[ReLU(z)=\left\{\begin{array}{ll}z&\text{if }z\ge 0\\0&otherwise\end{array}\right.\]
 and four input nodes. The trained network is represented by two matrices of  weights 
\begin{equation*}
w^I = 
\begin{pmatrix}
w_1^1 & w_2^1 & w_3^1 & w_4^1 \\ 
w_1^2 & w_2^2 & w_3^2 & w_4^2 
\end{pmatrix}
\hspace{5mm}\text{and}\hspace{5mm}
w^o=\begin{pmatrix}w_1^o&w_2^o\end{pmatrix}
\end{equation*}
for the input $n=w^II$ and the output $o=w^or$, respectively. The class decision for an input object $I=\begin{pmatrix}i_{1}&i_{2}&i_{3}&i_{4}\end{pmatrix}^t$ is given by
$$\hat{y}:=th_\tau(o)=th_\tau(\texttt{ANN}(I))=th_\tau(w^oReLU(w^II)).$$

\subsection*{\texttt{ANN} Partition}

The first step is to partition the input space $\mathbb{R}^n$ of an \texttt{ANN} into \emph{partition cells}. The key elements of the partition are the ReLU-nodes, which represent hyperplane separations of the input space. Let's demonstrate the partition with our example of a trained, simple \texttt{ANN}.  For any input object $I=\begin{pmatrix}i_{1}&i_{2}&i_{3}&i_{4}\end{pmatrix}^t$, each ReLU-node $r_i$ is either  \emph{inactive} ($r_i = 0$) or \emph{active} ($r_i > 0$), represented here by a binary code where 0 denotes inactive and 1 denotes active. With our two ReLU-nodes, we obtain one of four different cases $00, 01, 10, 11$ for every $I$, representing four partition cells.   Every inactive ReLU-node deactivates the mapping of its subtree below. On the other hand, every active ReLU-node represents  the identity function. As a result, we obtain a linear mapping for every partition cell $p$:
\begin{enumerate}
\item $p_{00}: o_{00} = 0\cdot i_1+0\cdot i_2+0\cdot i_3+0\cdot i_4$ for partition condition: $n_1\le 0\land n_2\le 0$
\item $p_{01}: o_{01} = (w^o_2w^2_1)\cdot i_{1}+(w^o_2w^2_2)\cdot i_{2}+(w^o_2w^2_3)\cdot i_{3}+(w^o_2w^2_4)\cdot i_{4}$ for partition condition: $n_1\le 0\land n_2> 0$
\item $p_{10}: o_{10} = (w^o_1w^1_1)\cdot i_{1}+(w^o_1w^1_2)\cdot i_{2}+(w^o_1w^1_3)\cdot i_{3}+(w^o_1w^1_4)\cdot i_{4}$ for partition condition: $n_1> 0\land n_2\le 0$
\item $p_{11}$: $o_{11}=o_{10}+o_{01}$ for partition condition:\\ $n_1> 0\land n_2> 0$
\end{enumerate}
Please note that a resulting  linear combination is not an attribute-wisely additive measure.  That is,   negative weights may occur.

\begin{figure}
\begin{center}
\includegraphics[scale=0.5]{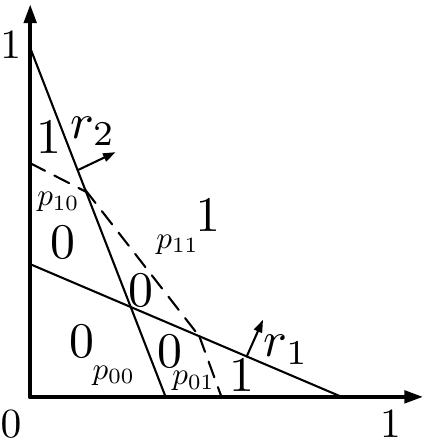}
\caption{\label{fig:relu} ReLU partition cells $p_{\_\_}$ based on hyperplanes and a class separation polyline in $[0,1]^2$}
\end{center}
\end{figure}

Figure~\ref{fig:relu} depicts the partition cells $p_{\_\_}$ of a trained \texttt{ANN} with  two ReLU-nodes and two input values from the range $[0,1]$ (in contrast to the four input values of our small example).   The ReLU-nodes $r_1$ and $r_2$ correspond to hyperplanes, which are drawn as solid lines, inducing the partitioning. The dashed line represents the class decision polyline defined by the threshold operator $th_\tau$ on the output score. All objects greater than the separation polyline are 1-objects\footnote{Objects with label 1}. The others are 0-objects.

\subsection*{Logical Interpretation of an \texttt{ANN} Partition Cell}

The second step of our approach is the logical interpretation of a partition cell. This requires interpreting the input value $x_i[j]\in\mathbb{R}$ of the $j$-th attribute of input object $i$ as the degree of fulfillment of an object property \cite{schmitt2006quantum,schmitt2008qql,schmitt2013logic}, akin to the membership value in  fuzzy set theory \cite{Zad65}.  For logical interpretation,   a map  of all attribute values to the unit interval $[0,1]$ is required. Thus,  we employ a monotonic increasing function $m_j()$, where $m_j(x_i[j])\in [0,1]$ for attribute $j=1,\ldots,n$ and all objects $i$.
The function $m_j()$, is application dependent and can be seen as  a fuzzification known from fuzzy set theory.


Furthermore, we require $2^n$ minterm values $mt_i[k]$ instead of $n$ values $m_j(x_i[j])$  as input for the network. Minterm values are conjunctions of negated or non-negated attribute values $m_j(x_i[j])$, expressing  interactions among attributes. In accordance with \cite{Sch22adbis,Sch22ideas}, we define
\begin{eqnarray}\label{eqn:minterm}
mt_i[k]:=\Pi_{j=1}^n \left(1-m_j(x_i[j])\right)^{1-b_j^k}\cdot m_j(x_i[j])^{b_j^k}
\end{eqnarray} 
where $k=0,\ldots,2^n-1$ is the minterm identifier and $b^k=b^k_1\ldots b^k_n$ is the   bit code of  $k$.
We denote minterm-based training data as $TR_{mt}:={(mt_i,y_i)}$ and assume an \texttt{ANN}  to be trained with $TR_{mt}$ and a loss function, e.g. MSE, with high accuracy.


For our small example and for an object $i$,  let $a$ represent the value of attribute $A$ and $b$ represent the value of attribute $B$,   that is, $a:=m_1(x_i[1])$ and $b:=m_2(x_i[2])$,  indicating the degree of fulfillment that the value is high. We construct the input $I$ using minterm values as:
\begin{eqnarray*}
I&=&\begin{pmatrix}i_{1}&i_{2}&i_{3}&i_{4}\end{pmatrix}^t:=\begin{pmatrix}[\overline{a}\overline{b}]&[\overline{a}b]&[a\overline{b}]&[ab]\end{pmatrix}^t\\
&=&\begin{pmatrix}mt_i[0]&mt_i[1]&mt_i[2]&mt_i[3]\end{pmatrix}^t.
\end{eqnarray*}
A conjunction corresponds to multiplication, and a negation corresponds to subtraction from 1, see Equation~\ref{eqn:minterm}.  For example, $i_2$ corresponds to  $\overline{a}b=\overline{a}\land b$ and is evaluated as $(1-a) \cdot b$.  

Each \texttt{ANN} partition cell determines  a linear combination of   weights $mw[k]$ on minterm values  $mt_i[k]$.  We will interpret $mw$ as  logic expressions.  The integer $k=0,\ldots,2^{n}-1$ identifies a minterm and its bits $b^k_j$ refer to attributes being negated (0) or non-negated (1).   A logic expression can be viewed as a pattern based on logic (interactions between attributes) for identifying 1-objects and distinguishing them from 0-objects. 

For example, let the linear combination of an \texttt{ANN} partition cell be $o=0.9\cdot i_1+0.4\cdot i_2+0.7\cdot i_3+0.8\cdot i_4$, see Table \ref{tab:minterm_weights}. 

Our core idea is to scale all minterm weights to the range $[0,1]$ and then to approximate every minterm weight  as a bit code $2^02^{-1}2^{-2}\ldots$ of a fixed  number of digits, see Table~\ref{tab:minterm_weights}. Every bit code level (column) induces a logic expression as a disjunction of minterms, see Table~\ref{tab:logic_expressions}. The logic expression at the $2^{-l}$ level has double the factor value compared to the next  level: $ 2^{-l}=2\cdot 2^{-(l+1)}$. The linear combination over the minterm values is approximated by the sum of the arithmetic evaluations over all derived logical expressions.

\begin{figure*}
\begin{center}
\includegraphics[scale=0.35]{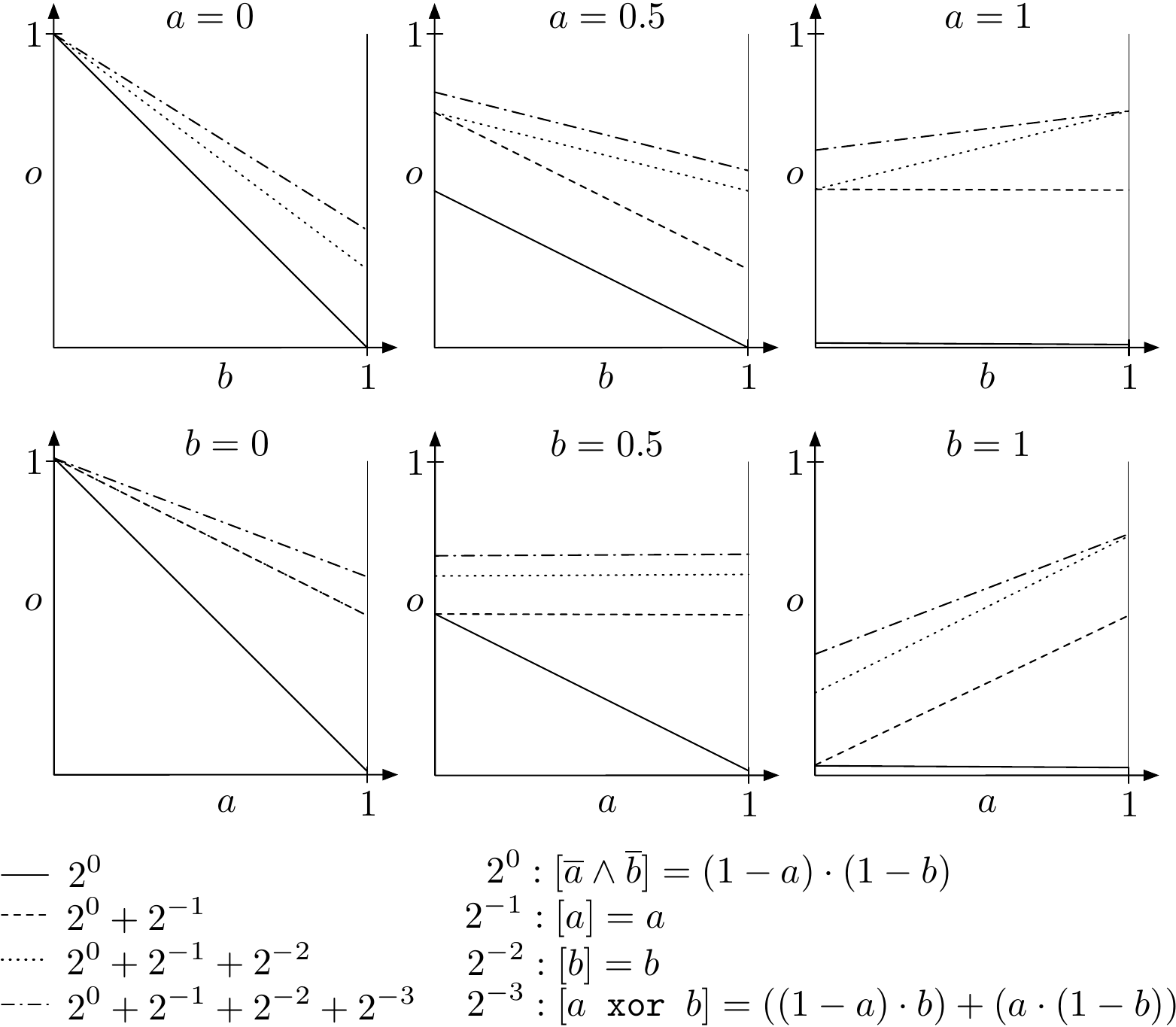}
\end{center}
\caption{\label{fig:trends} Trend diagrams  for evaluations of logic expressions of different bit code levels}
\end{figure*}

From the logic expressions on different bit code levels, we can readily derive a trend analysis (see Figure~\ref{fig:trends}) in order to investigate the interactions between $a$ and $b$ and see their effects on the output value $o$.

%

\begin{table*}
\caption{Example minterm weights and their bit codes
\label{tab:minterm_weights}}
\begin{center}
\begin{tabular}{c|c|c||cccc|c}
minterms& $k$ & $mw$ & \multicolumn{4}{|c|}{bit code of $mv$}&bit code\\
&  & & $2^0$ & $2^{-1}$ & $2^{-2}$ & $2^{-3}$& value \\\hline
$\overline{a}\overline{b}$ & $0$ & 0.9 & 1 & 0 & 0 & 0 & 1 \\
$\overline{a}b$ & $1$ & 0.4 & 0 & 0 & 1 & 1 & 0.375 \\
$a\overline{b}$ & $2$ & 0.7 & 0 & 1 & 0 & 1 & 0.625 \\
$ab$ & $3$ & 0.8 & 0 & 1 & 1 & 0 & 0.75 
\end{tabular}
\end{center}
\end{table*}


\begin{table*}
\caption{Weighted logic expressions\label{tab:logic_expressions}}
\begin{center}
\begin{tabular}{c||c|c|c}
level & active minterms & logic expression & arithm.  evaluation \\\hline
$2^0$& $\overline{a}\overline{b}$  &$\overline{a\text{ or } b}$ & $ 2^0\cdot(1-a)\cdot (1-b)$\\
$2^{-1}$& $a\overline{b},ab$ & $a$&$2^{-1}\cdot a$\\
$2^{-2}$&  $\overline{a}b,ab$& $b$&$2^{-2}\cdot b$\\
$2^{-3}$& $\overline{a}b,a\overline{b}$& $a\text{ xor }b$&$2^{-3}\cdot\left((1-a)\cdot b+a\cdot (1-b)\right)$ 
\end{tabular}
\end{center}
\end{table*}

After having given a brief overview of our method,  we will discuss it  with more details.

\section{\texttt{ANN} Partition Cells}
\label{sec:partitions}

Let's assume the trained network \texttt{ANN} has $l$ ReLU-nodes. We assign to every input object $mt_i$ the ReLU-status bit vector $relu_i\in\{0,1\}^l$, where $relu_i[m]$ represents the status of the $m$-th ReLU-node for $m=1,\ldots ,l$. A value of 1 (active node) for a ReLU-node indicates that its input value is non-negative, while 0 (inactive node) indicates a negative value. We interpret $relu_i$ as a bit code of an integer value $p_i\in {0,\ldots,2^l-1}$:
 $$p_i:=\sum_{m=1}^lrelu_i[m]\cdot 2^{l-m}.$$
We refer to the value $p_i$ of the $i$-th input object as its \emph{partition cell number}. Using equal partition cell numbers, we  decompose a set of input objects into equivalence classes, which we denote as \emph{partition cells}. Applied to $TR$, we obtain:
$$TR_{mt}=\bigcup_{p=0}^{2^l-1}\{(mt_i,y_i)|p_i=p\}.$$
All objects within the same partition cell share the same set of active or inactive ReLU-nodes.


For a given partition cell number $p$, we reduce the \texttt{ANN} to a \emph{network partition cell} denoted as \texttt{ANN}$_p$ by replacing all active ReLU-nodes with the constant factor 1 and inactive ReLU-nodes with the constant factor  0. As  result,  the ReLU-decisions do not have any effect  within a partition cell.   As the stacking of several linear maps produces again a  linear map, we can assign to every reduced network \texttt{ANN}$_p$ a linear map of the minterm values of an input object $i$ given that $p_i=p$:
\begin{equation}
\texttt{ANN}_p(mt_i)=\sum_{k=0}^{2^n-1} mw_p[k]*mt_i[k].\label{scalarproduct}
\end{equation}
The vector $mw_p\in\mathbb{R}^{2^n}$ represents the minterm weights of the linear combination of minterm values of input objects with partition cell number $p$. This vector $mw_p$ describes  the network partition cell $\texttt{ANN}_p$. As result,  a simple \texttt{ANN} with $l$ ReLU-nodes can be fully described by $2^l$ linear maps over minterm values.  

A minterm weight vector $mw_p$ may be directly used for computing a Shapley-value $Sh_i$  \cite{Rot88} in order to learn the impact of a single attribute   $a_i$ on the output:
$$Sh_i=\sum_{S\subseteq N\setminus \{i\}}\frac{(n-1-|S|)!|S|!}{n!}\left(v(S\cup \{i\})-v(S)\right).$$
The input for computing the Shapley values are the set of indices  $N:=\{1,\ldots,n\}$  for the object attributes $a_1,\ldots,a_n$ and the minterm values $v(S):=mw_p[k]$
where \mbox{$k=\sum_{i\in S}2^{i-1}$} is composed of  attributes that are regarded as non-negated attributes of minterm $k$.  Please note, that $v$ is in our case not necessarily increasing and $v(\emptyset)=mw_p[0]$ needs not to be  zero.   By summing up all Shapley values we obtain $$\sum_iSh_i=v(N)-v(\emptyset)=mw_p[2^n-1]-mw_p[0].$$
 For our example  in Table~\ref{tab:minterm_weights} with attributes $a,b$,  we obtain:
\begin{eqnarray*}
Sh_a=((mw_{a\overline{b}}-mw_{\overline{a}\overline{b}}) + (mw_{ab}-mw_{\overline{a}b}))/2 =0.1\\
Sh_b=((mw_{\overline{a}b}-mw_{\overline{a}\overline{b}}) + (mw_{ab}-mw_{a\overline{b}}))/2 =-0.2.
\end{eqnarray*}
Actually, it is not necessary to explicitly extract the minterm weights for every one of the $2^l$ partition cells. The minterm weights of a partition cell number  $p$ where $p=2^m$ for some integer $m$ represent the linear map of a cell with exactly one  active ReLU-node $m$.  Due to the linearity of the network layers,  we can readily derive the minterm weights of all possible partition cells by summing the minterm weights of the corresponding cells of single ReLU-nodes:
$$mw_p = mw_p^{b_1b_2\ldots b_l}=\sum_{i:b_i=1}mw_p^{2^{b_i}},$$
where $b_1b_2\ldots b_l$ is the bit code of partition cell number $p$.

\section{Logic Expressions}
\label{sec:cqqlp}

The evaluation of a  quantum-logic-inspired, binary CQQL\footnote{Commuting Quantum Query Language} classifier \cite{Sch22ideas} for a given object $o_i$ is defined as:
  \[
  cl^\tau_e(o_i) := th_\tau([e]^{o_i})\in\{0,1\},
  \]
where $e$ is a logic expression\footnote{Syntax is the same as propositional logic.} over $n$   input attributes $a_j$ and $[e]^{o_i}\in[0,1]$ is its arithmetic evaluation on the attribute values $m_j(o_i[j])\in [0,1]$.
See  for example following logic expression and its evaluation:
$$[a_1\land\neg a_2]^{o_i}:=m_1(o_i[1])\cdot(1-m_2(o_i[2])).$$ A CQQL expression and its evaluation obey the laws of Boolean algebra\footnote{Note that fuzzy logic does not obey the laws of Boolean algebra, e.g.\ $[a_1\land a_1]=[a_1]$ and $[a_1\land \overline{a}_1]=0$.}, although the attribute values are no Boolean values.   

From Boolean algebra we know that every logic expression $e$ can be expressed in complete disjunctive normal form, meaning every $e$ can be identified by a subset of the set of all  $2^n$ complete minterms. We denote the minterms being elements of the subset \emph{active} with status value 1 and those not being elements of  the subset \emph{inactive} with status value 0. As  result, every logic expression $e$ is identified by a bit vector $mw^e\in \{0,1\}^{2^n}$ over all $2^n$ minterm status values.  In terms of probability theory, all $2^n$ minterms are regarded as the total set of mutually \emph{exclusive} complex events and the scaled attribute values as probabilities  of probabilistically independent atomic events. To evaluate a logic expression $e$, along with the minterm values $mt_i$ of an input object $o_i$, we define the arithmetic evaluation as:
  \begin{equation}\label{cqql}
 [e]^{o_i}:=\sum_{k=0}^{2^n-1} mw^e[k]*mt_i[k]=\sum_{k:mw^e[k]=1}mt_i[k].
 \end{equation}
Here, we applied the law from probability theory stating that the disjunction of mutually exclusive events corresponds to the summation of their probabilities.

The evaluation formula of Equation~\ref{cqql} resembles the evaluation of a network partition cell shown in Equation~\ref{scalarproduct}. The main difference lies in the \emph{real} numbers of $mw_p$ contrasted with the \emph{binary} numbers of $mw^e$. Thus,  to interpret the linear combination as stated in Equation~\ref{scalarproduct} as a logic expression, we have  to map real numbers to binary numbers.

\section{Mapping Real-valued Minterm Weights to Binary Minterm Weights}

Equation~\ref{scalarproduct} corresponds to the scalar product between $mt_i$ and $mw_p$. From linear algebra, we know that the scalar product is linear in both of its arguments. As a next  step, we scale all minterm weights of all partition cells $mw_p[k]\in\mathbb{R}$ with $p=2^m$ for some integer $m$ to values from $[0,1]$ using a linear function. We select the linear function $f$ that is strictly increasing and maps the largest weight $max:=\max_{k,p=2^m}mw_p[k]$  to 1 and the smallest weight $min:=\min_{k,p=2^m}mw_p[k]$ to 0:
 $$mw_p[k]':=f(mw_p[k])\in[0,1]\text{ where }$$
 $$f(x):=\frac{x-min}{max-min}\footnote{If $max= min$ then we set $f(x)=1$.}.$$
This leads to the following equivalence:
$$\texttt{ann}_p'(mt_i):= \sum_{k=0}^{2^n-1} f(mw_p[k])*mt_i[k]=f(\texttt{ann}_p(mt_i)).$$
Furthermore, since $\sum_kmt_i[k]=1$ and $0\le mt_i[k]\le 1$, we obtain:
 $$\texttt{ann}_p'(mt_i)\in[0,1].$$
 Since $f$ is linear and strictly increasing the order relation of object evaluations is preserved for all $i_1,i_2,\tau$ and $\tau':=f(\tau)$:

\begin{eqnarray*}
\texttt{ann}_p(mt_{i_1})<\tau< \texttt{ann}_p(mt_{i_2})\Leftrightarrow\\ \texttt{ann}'_p(mt_{i_1})<\tau'< \texttt{ann}'_p(mt_{i_2}).
\end{eqnarray*}
As  result, scaling minterm weights to the unit interval does not modify the semantics of a classifier. In the following we assume normalized minterm weights and the new threshold $\tau'$.

As the next step, we approximate every minterm weight $mw_p[k]\in[0,1]$ of a partition cell $p=2^m$ by bit-values of different bit code levels $2^{-bcl}$:
$$mw_p[k]\approx \sum_{bcl=0}^{bcl_{max}}b[bcl,k,p]\cdot 2^{-bcl}$$
where $b[bcl,k,p]\in\{0,1\}$ is a three-dimensional bit tensor over bit code levels $bcl$, minterms $k$, and partition cells  $p$.
To minimize the  absolute error  between $mw_p[k]$ and its bit code,  the value $mw_p[k]$ should be suitably rounded  before deriving the bits. 

The average absolute error  of the bit-approximation of one minterm weight is not higher than $2^{-(bcl_{max}+1)}$.
Since $\sum_kmt_i[k]=1$, the same average absolute error holds also  between $[e]^{o_i}$ and the summarized bitwise $bcl$-evaluation of $e$ for an object $o_i$. 

The constant $bcl_{max}$ defines how close the minterm weights are approximated and is chosen as a trade-off between too many bits resulting in too many logic expressions and too high approximation error. As the average absolute error vanishes exponentially, the value of $bcl_{max}$ can be small.

For every bit code level $bcl$  and partition  cell $\texttt{ANN}_p$ we define the bit code vector $mw^e_{p,bcl}\in\{0,1\}^{2^n}$ as
$$mw^e_{p,bcl}[k]:=b[bcl,k,p]$$
and obtain in accordance to Equation~\ref{cqql} a logic expression $e_{p,bcl}$.   Thus, we   approximate $\texttt{ANN}_p(mt_i)$ as
$$\texttt{ANN}_p(mt_i)\approx \sum_{bcl=0}^{bcl_{max}}2^{-bcl}\cdot 
[e_{p,bcl}]^{o_i}$$
$$
[e_{p,bcl}]^{o_i}= 
\sum_{k:mw^e_{p,bcl}[k]=1} mt_i[k].
$$

\section{Interpretation  of a Logic Expression $e_{p,bcl}$}
\label{sec:interpreting}

\begin{figure}
\caption{Example training data for a logic expression $mw^e_{p,bcl}$ expressed over minterms\label{fig:qldt-example}: dashed lines in resulting QLDT denote negated evaluation;  leaves are class decisions;  tree represents logic expression $e_{p,blc}={a}_2\lor (\overline{a}_2\land a_1)={a}_2\lor {a}_1$ and $[e_{p,bcl}]^{o_i}=m_2(o_i[2])+((1-m_2(o_i[2]))*m_1(o_i[1]))$} 

\begin{center}
\begin{tabular}{c|cc|c}
& \multicolumn{2}{|c|}{$x$} & $y$\\
$k$ &\multicolumn{2}{|c|}{bit code} & $mw^e_{p,bcl}$\\
 & $a_1$ & $a_2$ &\\\hline
 0 & 0 & 0 & 0\\
 1 & 0 & 1 & 1\\
 2 & 1 & 0 & 1\\
 3 & 1 & 1 & 1
\end{tabular}
\hspace{10mm}
\begin{minipage}{20mm}
\includegraphics[scale=0.3]{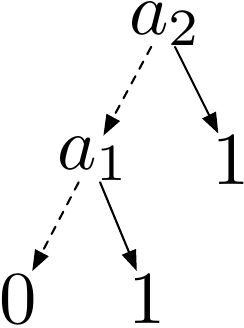}
\end{minipage}
\end{center}
\end{figure}

Of course, a bit vector $mw^e_{p,bcl}\in\{0,1\}^{2^n}$ is not an intuitive representation of a logic expression $e_{p,bcl}$. It simply tells us for every minterm whether it is active or not.  Therefore, we construct a quantum-logic-inspired decision tree (QLDT \cite{Sch22adbis}) from $mw^e_{p,bcl}$ and present it to the user for interpretation. The QLDT is just an alternative representation of $mw^e_{p,bcl}$. The main construction idea is to use the pairs $\{(\text{bit\_code}(k),mw^e_{p,bcl}[k])|k=0\ldots 2^n-1\}$ over the minterms $k$ as training data for learning a classical decision tree. The bit code levels of $k$ represent the original attributes being negated or non-negated. The attributes near the root of the resulting tree are more effective than those further away for distinguishing active from inactive minterms.   An example of a training data and resulting QLDT is given in Figure~\ref{fig:qldt-example}.  

Although a QLDT tree looks like a binary decision tree,  their evaluations differ. Every path to an active leaf corresponds to the product of its negated or non-negated attribute values. The    path evaluations to all active leaves are  summed up and checked against the final threshold in order to obtain the final decision. This evaluation can be understood as applying rules of probability theory to the probabilities of probabilistically independent events (attributes). Branches to the left and right correspond to exclusive disjunctions of compound (conjunctive) events.   Note, that as a special case, if the input attribute values are just 1 or 0 then the tree becomes a traditional decision tree.

Another benefit of exploiting a logical expression $e_{p,bcl}$ is the possibility to check it against a maybe given hypothetical logic expression $e_{hypo}$ from a user.   Questions are, for example, whether a hypothesis implies $e_{p,bcl}$  or vice versa,  or how strong their semantic overlap is. Such questions can be readily answered by comparing their corresponding  bit vectors $mw^{e}_{p,bcl}$ and $mw^{e_{hypo}}$ based on their minterm values.  If we regard the bit vectors as sets of active minterms $active(mw^e)=\{k|mw^e[k]=1\}$, then we  compute:
\begin{eqnarray*}
v_{11}&:=&|active(mw^{e}_{p,bcl})\cap active(mw^{e_{hypo}})|\\
v_{10}&:=&|active(mw^{e}_{p,bcl})\cap active(mw^{\neg e_{hypo}})|\\
v_{01}&:=&|active(mw^{\neg e}_{p,bcl})\cap active(mw^{e_{hypo}})|\\
v_{00}&:=&|active(mw^{\neg e}_{p,bcl})\cap active(mw^{\neg e_{hypo}})|
\end{eqnarray*}
where $v_{11}+v_{10}+v_{01}+v_{00}=2^n$. Based on these values we can compute measures like accuracy, precision and recall. For example, accuracy is given by $acc:=(v_{11}+v_{00})/{2^n}$.  A value $acc=1$ means that $e$ and $e_{hypo}$ share the same active minterms and are therefore identical.  An implication refers to the case $v_{10}=0$  or $v_{01}=0$ where the sets of active minterms are in a subset relation.

\section{Experimental Study}
\label{sec:study}

We apply our approach to the binary classification problem of the banknote authentication dataset\footnote{\url{https://github.com/topics/banknote-authentication-dataset}}. This dataset contains wavelet-processed image data, which are used to determine the authenticity of banknotes. The training data include values for the four attributes: variance (v), skewness (s), curtosis (c), and entropy (e), as well as target information indicating whether the banknote is authentic or not. The dataset contains 1220 objects with known target values. We trained an \texttt{ANN} consisting of $2^4$ input nodes representing minterms, three hidden ReLU-nodes $r_i$, and one output node with a carefully chosen threshold. As a result, the trained \texttt{ANN} demonstrates an accuracy of 100\%.

Checking all training data against ReLU-conditions yields four  non-empty  network partition cells, as shown in Table~\ref{tab:banknote_partitions}. Column '$B$' represents the number of authentic banknotes, and '$\overline{B}$' represents the number of non-authentic banknotes contained in the respective partition cell.   By examining the effect of the ReLU-nodes $r_i$ on authentication, we observe that  $r_2$  separates the authentic banknotes from the non-authentic ones at best. Therefore, we focus exclusively on partition cell $\texttt{ANN}_2$ since it refers exclusively to $r_2$.

\begin{table}
\caption{Banknote authentication: network partition cells\label{tab:banknote_partitions}}
\begin{center}
\begin{tabular}{c||c|c|c||c|c}
Partition cell & $r_1$ & $r_2$ & $r_3$ & $B$ & $\overline{B}$\\\hline
$\texttt{ANN}_3$ & 0 & 1 & 1 & 513 & 0\\
$\texttt{ANN}_0$ & 0 & 0 & 0 & 0 & 609\\
$\texttt{ANN}_7$ & 1 & 1 & 1 & 95 & 0\\
$\texttt{ANN}_2$ & 0 & 1 & 0 & 2 & 1
\end{tabular}
\end{center}
\end{table} 

Table~\ref{tab:banknote_partition_two} shows the minterm weights of partition cell  $\texttt{ANN}_2$  after $[0,1]$-scaling and their approximated bit codes. Each bit code level (column) represents a set of active minterms, which is regarded as a logical expression. 
The weight sum $9.46$  represents the total energy\footnote{Energy means here the contribution of the corresponding  minterm weights to the network output.} of the given minterm weights and is used to calculate the relative energy ($E$ in percent) of the column-wise logical expressions. For example, the relative energy of $bcl=1$ with 11 set bits is 
approximately $58\%$,  calculated as $2^{-1} \cdot 11 / 9.46 \cdot 100\%$. 
The last column shows the approximated minterm weights as the sum of bits. The approximation error between the original minterm weights and their approximated bit code level representations is small.

\begin{table*}
\caption{Banknote authentication: minterm weights of network partition cell $\texttt{ANN}_2$ and their bit codes\label{tab:banknote_partition_two}}
\begin{center}
\begin{tabular}{c|cccc||c|cccc|c}
$k$ & $v$ & $s$ & $c$ & $e$ & Minterm weight & $2^0$ &  $2^{-1}$ &  $2^{-2}$ &  $2^{-3}$ &  $\Sigma$\\\hline
0 & 0 & 0 & 0 & 0 & 1 			& 1 & 0 & 0 & 0 & 1\\
1 & 0 & 0 & 0 & 1 & 0.918  	& 0 & 1 & 1 & 1 & 0.875\\
2 & 0 & 0 & 1 & 0 & 0.688 	& 0 & 1 & 1 & 0 & 0.75\\
3 & 0 & 0 & 1 & 1 &  0.751 	& 0 & 1 & 1 & 0 & 0.75\\
4 & 0 & 1 & 0 & 0 &  0.625 	& 0 & 1 & 0 & 1 & 0.625\\
5 & 0 & 1 & 0 & 1 &  0.546 	& 0 & 1 & 0 & 0 & 0.5\\
6 & 0 & 1 & 1 & 0 &  0.660 	& 0 & 1 & 0 & 1 & 0.625\\
7 & 0 & 1 & 1 & 1 &  0.431 	& 0 & 0 & 1 & 1 & 0.375\\
8 & 1 & 0 & 0 & 0 & 0.783 	& 0 & 1 & 1 & 0 & 0.75\\
9 & 1 & 0 & 0 & 1 & 0.731 	& 0 & 1 & 1 & 0 & 0.75\\
10 & 1 & 0 & 1 & 0 & 0.291 	& 0 & 0 & 1 & 0 & 0.25\\
11 & 1 & 0 & 1 & 1 &  0.635 	& 0 & 1 & 0 & 1 & 0.625\\
12 & 1 & 1 & 0 & 0 &  0.525 	& 0 & 1 & 0 & 0 & 0.5\\
13 & 1 & 1 & 0 & 1 &  0.613 	& 0 & 1 & 0 & 1 & 0.625\\
14 & 1 & 1 & 1 & 0 &  0 			& 0 & 0 & 0 & 0 & 0\\
15 & 1 & 1 & 1 & 1 &  0.259 	& 0 & 0 & 1 & 0 & 0.25\\\hline
$\Sigma$ &&&&& 9.46 & $1/{2^0}$ & $11/{2^1}$ & $8/{2^2}$ & $6/{2^3}$ & 9.25\\
$E$ &&&&& 100\% & 11\% & 58\%  & 21\% & 7\% & 97 \%
\end{tabular}
\end{center}
\end{table*} 

Next, we determine how many bit code levels we actually need for a good prediction. If we  consider  only $bcl\in \{0\}$  with an energy of $11\%$,  we achieve a classification accuracy  of $78\%$ by evaluating $mw^e_{2,0}$.  For combined levels $bcl\in\{0,1\}$ (energy $11\%+58\%=69\%$),   i.e.   with $mw^e_{2,0}+mw^e_{2,1}$, we achieve an accuracy of $84\%$, and for levels $bcl\in\{0,1,2\}$ (energy $11\%+58\%+21\%=90\%$), we achieve an accuracy of $98\%$.   Due to the high accuracy value of $98\%$,  we focus on levels up to level $2$ and ignore $mw^e_{2,3}$.

Let us now examine the logical expressions for the single bit code levels. The bit code level $bcl=0$ yields just one minterm $\overline{v}\land \overline{s}\land \overline{c}\land \overline{e}$, indicating that if all attribute values are low, we obtain a positive authentication.  

\begin{figure*}
\begin{center}
\includegraphics[scale=0.45]{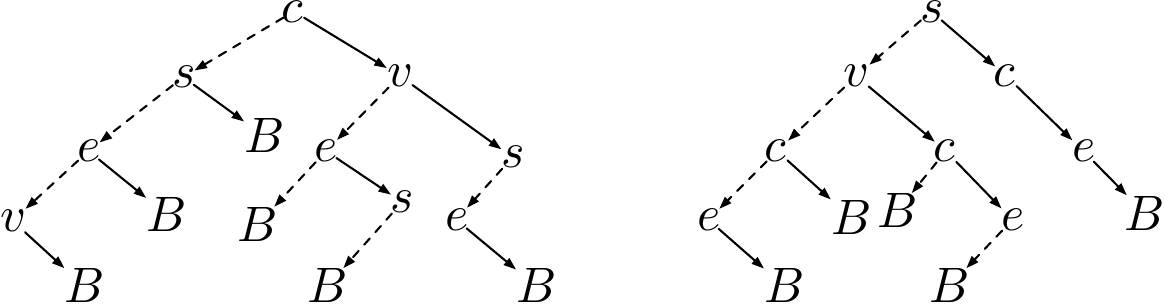}
\caption{\label{fig:tree1} Logic expressions of bit code level $bcl=1$ (left) and $bcl=2$ (right) as trees; dashed line for negated evaluation and solid line for non-negated evaluation;  a leaf $B$ stands for an active minterm for the banknote authentication; inactive minterms are not shown; $v,s,c,e$ are short notations of the input object attributes}
\end{center}
\end{figure*}

An intuitive representation of a logic expression is a binary tree, where the left child corresponds to a low attribute value and the right child to a high value. We construct the binary tree from the minterm information by applying a decision tree algorithm following the QLDT-approach \cite{Sch22adbis}, as shown in Figure~\ref{fig:tree1}. 
For example, the rightmost branch to $B$ means that banknote authentication is supported if skewness, curtosis, and entropy are high.

If we want to restrict our classification problem to the attributes $v$ and $s$ then we  derive their $vs$-minterm $00$ value by adding the original  values of the minterms $0000,0001,0010,0011$. Analogously, we obtain the values for the $vs$-minterms $01$, $10$, and $11$. After rescaling, we obtain bit code values as shown in Table~\ref{tab:banknote_partition_two_vs}. The logical interpretation yields '$\overline{v}\land\overline{s}$' with factor $2^0$ and '$v \texttt{ xor } s$' with factor $2^{-1}$,  which  tells us the character of interaction between $v$ and $s$. Note that these expressions match  well the trees in Figure~\ref{fig:tree1}. Of course, dropping the attributes $c$ and $e$ means a loss of information. However, the classification accuracy of $87\%$ relying solely on $v$ and $s$ is surprisingly high.

\begin{table*}
\caption{Banknote: minterm weights of partition cell  $\texttt{ANN}_2$ projected to attributes $v$ and $s$\label{tab:banknote_partition_two_vs}}
\begin{center}
\begin{tabular}{c|cc||c|cccc|c}
Minterm & $v$ & $s$ & Minterm weight & $2^0$ &  $2^{-1}$ &  $2^{-2}$ &  $2^{-3}$ &  $\Sigma$\\\hline
0 & 0 & 0 & 1 			& 1 & 0 & 0 & 0 & 1\\
1 & 0 & 1 & 0.44			& 0 & 1 & 0 & 0 & 0.5\\
2 & 1 & 0 & 0.53 			& 0 & 1 & 0 & 0 & 0.5\\
3 & 1 & 1 & 0 			& 0 & 0 & 0 & 0 & 0\\\hline
$\Sigma$ &&& 1.97 & $1/{2^0}$ & $2/{2^1}$ & $0/{2^2}$ & $0/{2^3}$ & 2\\
$E$ &&& 100\% & 51\% & 51\%  & 0\% & 0\% & 102\% 
\end{tabular}
\end{center}
\end{table*} 

%
%
%
%

\section{Conclusion}

We addressed the problem of interpreting simple  \texttt{ANN}s using logic expressions by constructing a bridge between them.  Logic expressions are  visualized as binary logic trees (QLDT).

ReLU-nodes of an \texttt{ANN} introduce non-linearity to the network, which is necessary for solving non-linear classification tasks. Our approach is to partition a trained simple \texttt{ANN} by use of ReLU-conditions in order to obtain linear maps for every partition cell. The minterm weights of a linear map are interpreted as bit numbers. Every bit code level  of a partition cell  defines a set of minterms and corresponds to a logic expression.

The main benefit of our approach is to interpret an \texttt{ANN} by means of logic. Logic provides a powerful toolset for analyzing how attributes interact with each other.

Our approach relies on using minterm values as network input.  Since there are $2^n$ minterms for $n$ object attributes, the number $n$ is restricted  to be small.
In future work, we will strive to adapt this approach to more complex neural networks.

\bibliographystyle{apalike}
{\small
\bibliography{paper}}

\end{document}